\documentclass[final,3p,times,sort&compress]{elsarticle}
\usepackage{color}
\usepackage{url}
\usepackage{multirow,setspace,times,amssymb,amsmath,graphicx,color,subfigure,url}
\usepackage{dsfont}
\usepackage{threeparttable}
\usepackage{rotating}

\usepackage{dcolumn}
\usepackage{listings}

\journal{Elsevier}

\begin{document}

\begin{frontmatter}

\title{How does economic policy uncertainty comove with stock markets: New evidence from symmetric thermal optimal path method}

\author[SUIBE]{Ying-Hui Shao}
\ead{yinghuishao@126.com}

\author[SILC]{Yan-Hong Yang \corref{cor1}}
\ead{yanhongyang.ecust@foxmail.com}

\author[SoB,RCE,DoM]{Wei-Xing Zhou \corref{cor1}}
\ead{wxzhou@ecust.edu.cn}

\cortext[cor1]{Corresponding authors.}

\address[SUIBE]{School of Statistics and Information, Shanghai University of International Business and Economics, Shanghai 201620, China}
\address[SILC]{SILC Business School, Shanghai University, Shanghai 201899, China}
\address[SoB]{
	School of Business, East China University of Science and Technology, Shanghai 200237, China}
\address[RCE]{Research Center for Econophysics, East China University of Science and Technology, Shanghai 200237, China}
\address[DoM]{School of Mathematics, East China University of Science and Technology, Shanghai 200237, China}

\begin{abstract}
 We revisit the dynamic relationship between domestic economic policy uncertainty and stock markets using the symmetric thermal optimal path (TOPS) method. We observe different interaction patterns in emerging and developed markets. Economic policy uncertainty drives the stock market in China, while stock markets play a leading role in the UK and the US. Meanwhile, the lead-lag relationship of the three countries reacts significantly to extreme events. Our findings have important implications for investors and policy makers. 
\end{abstract}
\end{frontmatter}

\section{Introduction}
\label{s:introduction}
The interaction between economic policy uncertainty (EPU) and stock market has received wide attention in recent years, including the spillover effect 
\citep{he2020asymmetric}, the predictive power of policy uncertainty on stock markets \citep{bekiros2016economic,liu2017can,phan2018can}, the effects of policy uncertainty on stock market systematic
risk \citep{tsai2017source}. A plethora of studies reveal the negative effect of EPU on stock markets \citep{kang2013oil,ko2015international,yang2016dynamic, christou2017economic, guo2018asymmetric,he2020asymmetric,liang2020us}. An increase in the US EPU causes a decrease in the US stock returns \citep{arouri2016economic}. Policy uncertainty shocks also negatively affect the US stock-bond correlations \citep{li2015economic}. The absolute changes in the US EPU have a negative influence on the comovement of the Chinese and US stock markets \citep{li2017us}. Moreover, EPU negatively  predicts stock market returns at various horizons \citep{chen2017economic}. 

Furthermore, the linkage between policy uncertainty and the stock market is country-dependent \citep{phan2018can} and dynamic \citep{ko2015international}. 
Moreover, it is sensitive to market conditions \citep{you2017oil} and influential events like the 2008 financial crisis \citep{antonakakis2013dynamic,xiong2018time}. Compared with developed markets, emerging stock markets are less vulnerable to both domestic and international policy uncertainty \citep{das2018international}. 

While the current literature on the effect of stock markets on EPU is limited, several studies have reported that policy makers might react to stock market movements and adjust policy accordingly \citep{li2016causal}.
Hence stock markets can also have an impact on policy uncertainty \citep{antonakakis2013dynamic,li2016causal,yang2016dynamic}. An increase in stock market volatility raises policy uncertainty in the US \citep{antonakakis2013dynamic}. 
Other scholars also report a bidirectional causal relationship between EPU and stock returns \citep{li2016causal,dakhlaoui2016interactive,li2020does}.

So far, the existing studies mainly focus on the unidirectional impacts of international policy uncertainty on domestic stock markets at a monthly frequency. Moreover, a few researchers examine the difference in this causality between developing and developed markets. To give a more detailed analysis, we revisit the time-varying lead-lag structure between the two variables of China, the UK and the US using the symmetric thermal optimal path (TOPS) method \citep{sornette2005non,Zhou-Sornette-2006-JMe,Meng2016}. We employ country-specific empirical proxies constructed by \cite{baker2016measuring} and \cite{huang2020measuring} to measure economic policy uncertainty in China, the UK, and the US.

We contribute to current literature by providing dynamic evidence of the domestic EPU-stock market comovement via the TOPS method from daily data. The TOPS method enables a time-varying investigation and does not require the stationarity of signals. As a nonlinear and nonparametric technique, it has been fruitfully applied to studies of lead-lag dependency 
\citep{guo2011us,xu2017time,Guo2017,wang2017lead,shao2019time,yang2020time,yang2020times}. Furthermore, the findings of this research provides insights for various EPU-stock interaction patterns and market efficiency in developed and developing economies. Our findings may provide important implications for financial stability, risk management and stock return predictability.

The reminder of this paper is organized as follows. Section~\ref{S2:Data description} depicts data and summary statistics. Section~\ref{S3:Methodology} describes the methodology. Section~\ref{S4:Results} presents the dynamic lead-lag relationship between EPU and the stock market, and section~\ref{S5:conclusion} summarizes the paper.

\section{Data description}
\label{S2:Data description}

We consider the economies of China, the UK and the US for which the daily EPU data are available. The policy uncertainty series for the UK and the US come from \cite{baker2016measuring}, whose method is commonly used to measure real economic policy uncertainty. Considering data availability, we focus on China's policy uncertainty index developed by \cite{huang2020measuring}, which is based on work of \cite{baker2016measuring}. The news-based EPU dataset covers the period from 2000 to 2020. Meanwhile, we utilize the daily closing prices of the Shanghai Stock Exchange Composite index (SSEC) for China, the Financial Times Stock Exchange 100 index (FTSE 100) for the UK, and the Standard and Poor's 500 Index (S\&P 500) for the US.

\begin{table}[!htp]\addtolength{\tabcolsep}{2pt}
	\footnotesize
	\caption{Summary statistics of EPU and stock index returns.}
	\label{Tb:EPU:STOCK:Statistical:Summary:Daily}
	\medskip
	\centering
	\begin{threeparttable}
		\begin{tabular}{lllllll}
			\hline\hline
			& CNEPU & UKEPU & USEPU &SSEC&FTSE100&S\&P500  \\
			\hline			
			Mean&0.4635&0.5214&0.4948&0.5065&0.5509&0.5393\\
			Maximum&1.0000&1.0000&1.0000&1.0000&1.0000&1.0000\\
			Minimum&0.0000&0.0000&0.0000&0.0000&0.0000&0.0000\\
			Std.Dev&0.1086&0.0653&0.0787&0.0894&0.0589&0.0554\\
			Skewness&0.0992&-0.0207&0.0359&-0.6409&-0.3705&-0.5536\\
			Kurtosis&4.2635&5.8831&5.0753&7.6649&12.4175&15.9152\\
			$\rm JB_{\it p-\rm value}$\tnote{a} &0.0010&0.0010&0.0010&0.0010&0.0010&0.0010\\
			$\rm ADF_{\it p-\rm value}$\tnote{b}&0.0010&0.0010&0.0010&0.0010&0.0010&0.0010\\
			\hline\hline
		\end{tabular}
		\begin{tablenotes}[para,flushleft]
			Notes: \\
			\item[a]  $\rm JB$ is the Jarque-Bera test of normality, which is distributed as $\chi^2(2)$, and the $\rm JB_{\it p-\rm value}$ is the associated $p$-value. \\
			\item[b]  ADF is the Augmented Dickey-Fuller test of unit root, 
			the $\rm ADF_{\it p-\rm value}$ is the associated $p$-value. \\
		\end{tablenotes}
	\end{threeparttable}
\end{table}

\begin{figure}[!htp]
	\centerline{\includegraphics[width=16cm]{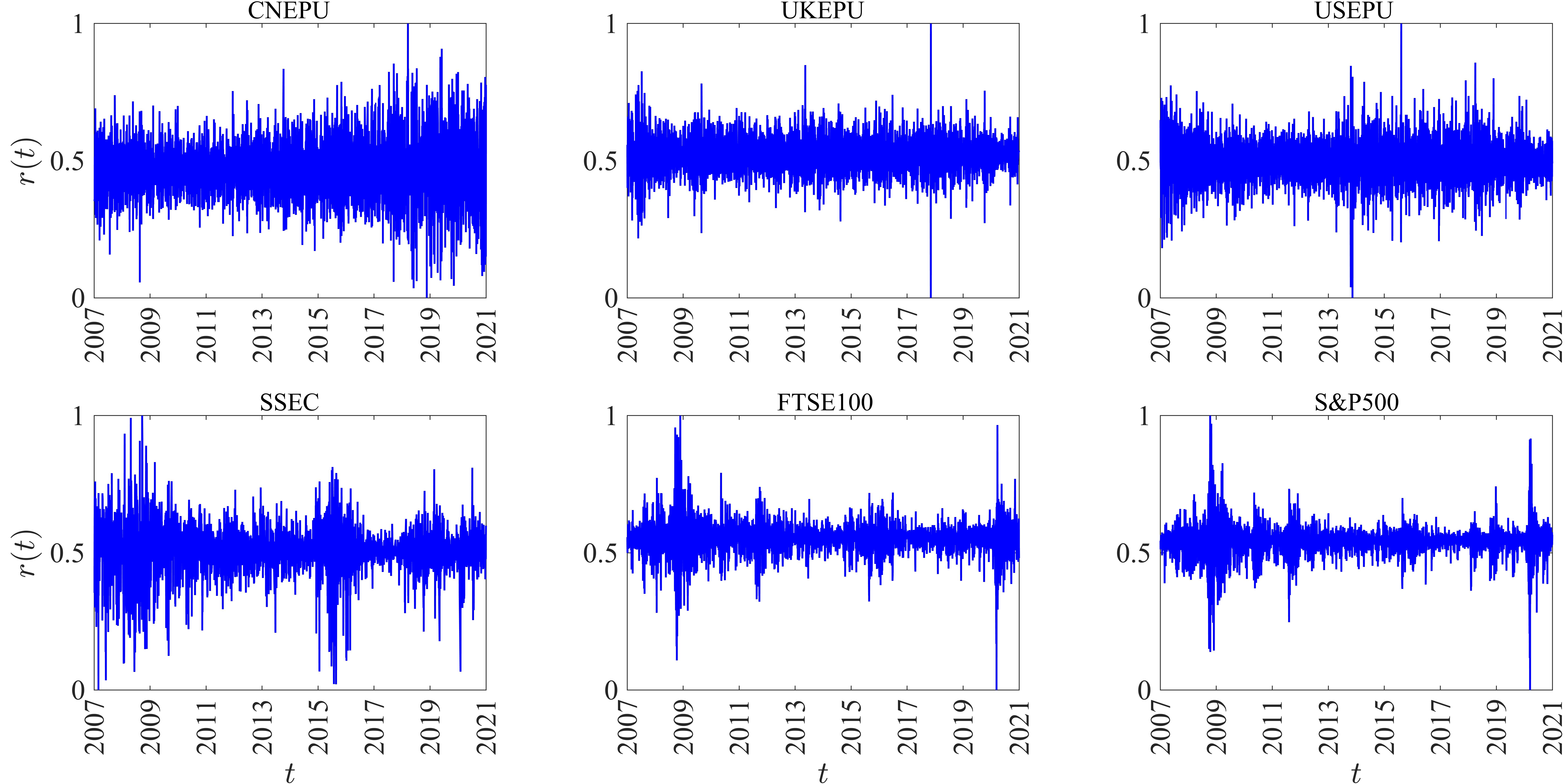}}
	\caption{EPU and stock market index returns.}
	\label{Fig:TOPS:Return:Xav:Daily}
\end{figure}

We take logarithmic returns for analysis. 
Fig.~\ref{Fig:TOPS:Return:Xav:Daily} and Table~\ref{Tb:EPU:STOCK:Statistical:Summary:Daily} illustrate the data. China's EPU has the largest standard deviation while the S\&P 500 has the lowest. All returns are skewed and heavy-tailed. The Jarque–Bera test also suggests that all variables are not normally distributed. The $p$-value of the augmented Dickey–Fuller (ADF) test does not transcend 0.05, indicating that the dataset is stationary. 

As illustrated in Fig.~\ref{Fig:TOPS:Return:Xav:Daily}, China's stock market went up and down in 2008, 2015 and 2019. The UK has witnessed wild swings in 2008 and 2020. The same phenomenon was also observed in the US stock market. Moreover, the UK's EPU showed similar changes as the US' EPU. 

\section{Methodology}
\label{S3:Methodology}
The TOPS method was introduced to effectively identify the structural changes in lead-lag relationships between two time series \cite{Meng2016}. The details of this method are depicted as follows.

Consider there are two standardized time series $X(t_1):t_1=0,\cdots,n-1$ and $Y(t_2):t_2=0,\cdots,n-1$, in which $X(t_1)$ and $Y(t_2)$ are the logarithmic returns of EPU and the stock index, respectively. First, we form a distance matrix $E_{X,Y}$ that allows us to compare the disparities between all the values of $X(t_1)$ with those of $Y(t_2)$ along the two time axes $t_1$ and $t_2$. The elements of the distance matrix $E_{X,Y}$ are defined as
\begin{equation}
\epsilon(t_1,t_2) = |X(t_1)-Y(t_2)|~.
\label{Eq:DistMatrix}
\end{equation}
The $n\times n$ distance matrix $E_{X,Y}$ further defines the mapping $t_1 \rightarrow t_2 = \phi(t_1)$ expressed as
\begin{equation}
\phi(t_1)=\min_{t_2}\{\epsilon(t_1,t_2)\},
\label{Eq:LocalMinimization}
\end{equation}
where Eq.~(\ref{Eq:LocalMinimization}) is a local minimization. To eliminate unreasonably large jumps or contradicting causality, \cite{sornette2005non} replace the local minimization in  Eq.~(\ref{Eq:LocalMinimization}) with the following global minimization
\begin{equation}
\min \limits_{\{\phi(t_1), t_1=0,1,\ldots,n-1\}} E:= \sum_{t_1=0}^{n-1}|X(t_1)-Y(\phi(t_1))|,
\label{Eq:GlobalMinimization}
\end{equation}
with a continuity constraint
\begin{equation}
0 \leq \phi(t_1+1)-\phi(t_1) \leq 1~.
\label{Eq:ContinuousConstraint}
\end{equation}
The continuous time limit of condition Eq.~(\ref{Eq:ContinuousConstraint}) is to ensure continuity in the one-to-one mapping $t_1 \rightarrow t_2 = \phi(t_1)$.

To solve the global optimization problem Eq.~(\ref{Eq:GlobalMinimization}) more efficiently, one can transform the original coordinates $(t_1,t_2)$ to $(t,x)$ as follows \cite{sornette2005non}
\begin{equation}
\left\{
\begin{array}{ccl}
t = t_2+t_1\\
x = t_2-t_1.
\end{array}
\right.
\label{Eq:AxesTransform}
\end{equation}
Then, the optimal thermal path $\langle x(t) \rangle$ of the TOPS method is determined as \cite{Meng2016}
\begin{equation}
\langle x(t) \rangle = \sum_{x}x\frac{\overrightarrow{G}(t,x)/\overrightarrow{G}(t) + \overleftarrow{G}(t,x)/\overleftarrow{G}(t)}{2}~,
\label{Eq:ThermalAveragePath TOPS}
\end{equation}
where $\overrightarrow{G}(t,x)$ is the local weight factor for the searching direction from past to future. And
\begin{equation}
\overrightarrow G(t) = \sum_{x} \overrightarrow G(t,x).
\end{equation}
Here, $\overrightarrow G(t,x)/\overrightarrow G(t)$ is the probability that a path will be at position $x$ at time $t$. To keep the direction of time required
by `causality', a feasible path arriving at $(t_1+1,t_2+1)$ can stem from $(t_1+1,t_2)$ vertically, $(t_1,t_2+1)$ horizontally, or $(t_1,t_2)$ diagonally. Thus, the local weights at $(t,x)$ can be calculated in a recursive manner via the following equation
\begin{equation}
\overrightarrow G(t+1,x)=[\overrightarrow G(t,x-1)+\overrightarrow G(t,x+1)+\overrightarrow G(t-1,x)]e^{-\epsilon(t+1,x)/T},
\label{Eq:Recursive:W}
\end{equation}
where $T$ is a parameter controlling the effect of noise.
Correspondingly, $\overleftarrow{}$ refers  the recursive process which is along the time-reversed direction.

Finally, one can unveil the dependence structure between EPU and the stock market by the value of $\langle x(t) \rangle$. EPU leads the stock market if $\langle x(t) \rangle>0$. Otherwise EPU lags the stock market when $\langle x(t) \rangle<0$. In particular, neither EPU nor the stock market is dominant when $\langle x(t) \rangle=0$.

\section{Results}
\label{S4:Results}

\begin{figure}[!h]
	\centerline{\includegraphics[width=16cm]{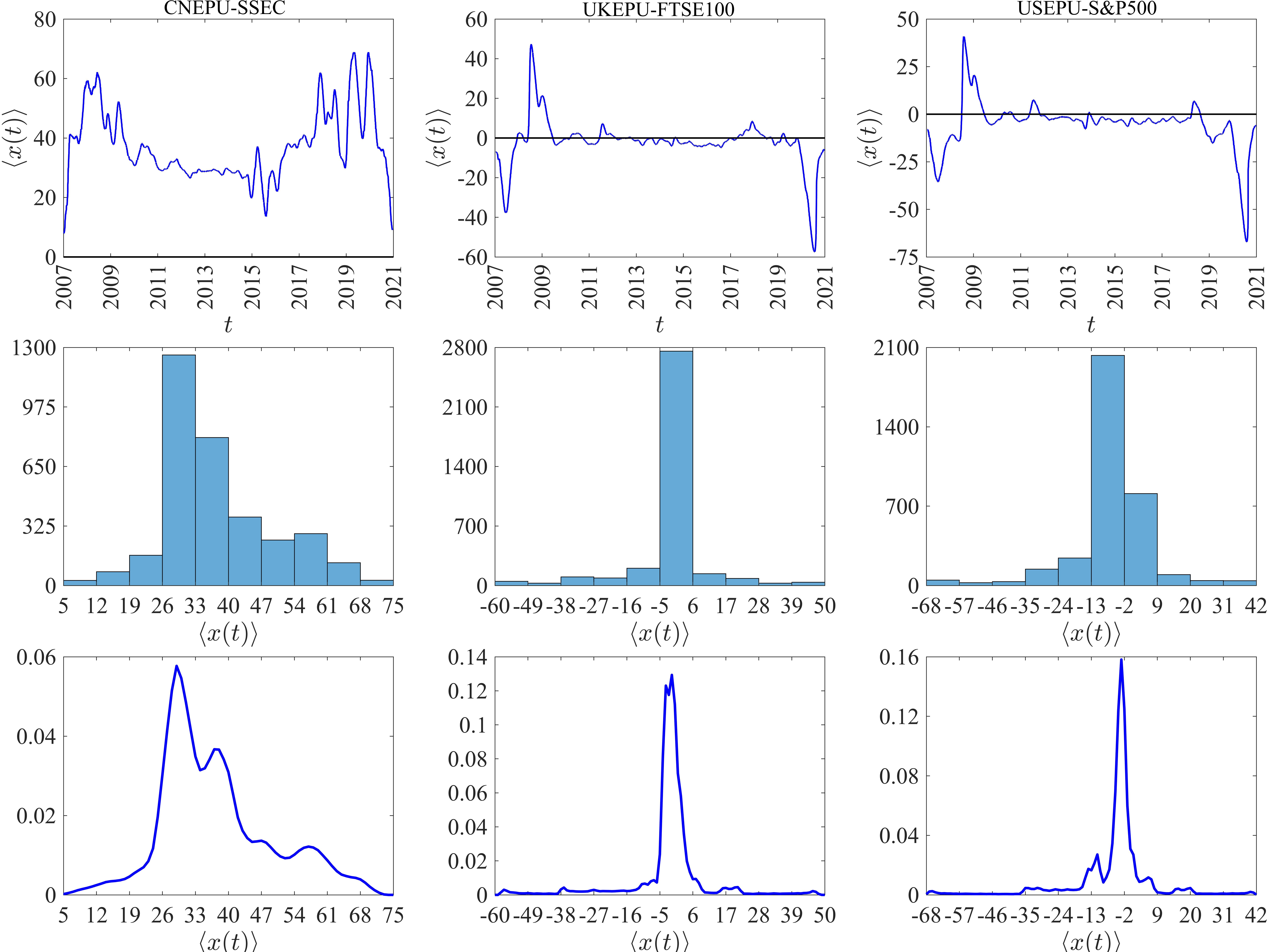}}
	\caption{Lead-lag path $\left\langle x(t)\right\rangle $, its histograms and probability density curves. 		
		From left to right, each column corresponds to $\left\langle x(t)\right\rangle $ of ``CNEPU vs. SSEC'', ``UKEPU vs. FTSE100'' and ``USEPU vs. S\&P500''.}
	\label{Fig:TOPS:Sta}
\end{figure}

Following \cite{Meng2016}, we examine the dynamic coherence between EPU and the stock market based on the TOPS method with $T$ = 2. 
We illustrate results in Fig.~\ref{Fig:TOPS:Sta}
and Table~\ref{Tb:xav:Statistical}. 

As shown in Fig.~\ref{Fig:TOPS:Sta} Chinese EPU dominates the stock market, which means the stock market in China is significantly influenced by policy uncertainty. Our findings are consistent with previous studies of \cite{guo2013monetary}. Their results suggest that the Chinese stock market is a ``policy market''.  Throughout the global financial crisis, the Chinese $\left\langle x(t)\right\rangle$ 
increases steadily. Obviously, $\left\langle x(t)\right\rangle$ peaks during the financial crisis in late 2008. Since 2009 it shows a clear downward trend, although there are periods of slight fluctuations. Then $\left\langle x(t)\right\rangle$ fluctuates around 30 until the Chinese stock market crash in 2015. During the trade disputes between China and the US since 2018, there is a general uptrend in $\left\langle x(t)\right\rangle$ with volatile movement. As China is highly export-dependent and the US is China's most important trading partner, it is not surprising that economic activity in China is affected by trade protection in the US \citep{tsai2017source}. There is an upsurge in anxiety and uncertainty about future trade policy in China during this period. 
However, since the initial outbreak of COVID-19 in December 2019 $\left\langle x(t)\right\rangle$ drops very sharply. 

The lead-lag path $\left\langle x(t)\right\rangle $ of China has 100.00\% positive values, which implies that EPU leads the stock market in China during the whole sample period. The mean value of Chinese $\left\langle x(t)\right\rangle $ is 37.31, and its median is 35.16, which also supports the view that EPU leads the stock market. 
As illustrated in Fig.~\ref{Fig:TOPS:Sta}, the lead-lag paths for China are mostly in the range of 26 to 40, indicating that the stock index lags EPU by 26 to 40 days. 

Fig.~\ref{Fig:TOPS:Sta} shows that there is a distinct lead-lag relationship between economic policy uncertainty and the stock market for a developed county, which is in line with work of \cite{das2018international}. Overall, the UK stock market index leads its EPU. The results provide evidence that this market is information efficient. From 2007, the UK $\left\langle x(t)\right\rangle $ decreases and then increases. Before 2008 the path is mostly negative, which means that the stock market leads EPU. Since the second half of 2008, positive $\left\langle x(t)\right\rangle $ is observed. The path increases continually and reaches its peak at 47.36 on July 16, 2008, which corresponds to the violent fluctuations in EPU. 
During this period, the lead-lag relationship reverses and EPU  leads the stock market. Then the lead-lag path $\left\langle x(t)\right\rangle $ decreases steadily until late 2009. Since then to 2020, there exists a slightly negative lead-lag structure in UK with occasional exceptions, which implies that the stock market takes the lead. After the COVID-19 outbreak in December of 2019, $\left\langle x(t)\right\rangle $ drops sharply to a minimum of -57.48 on July 27, 2020. 

As illustrated in Fig.~\ref{Fig:TOPS:Sta}, the probability density curve of the UK lead-lag path is almost symmetrical around $-2$. Table~\ref{Tb:xav:Statistical} documents the mean and median of the UK's $\left\langle x(t)\right\rangle $ at -2.21 and -1.27, respectively. Negative $\left\langle x(t)\right\rangle $ is observed on a higher percentage of the whole path~(69.15\%), which also suggests that the UK stock market movement precedes changes in the EPU. 

What stands out in Fig.~\ref{Fig:TOPS:Sta} is that the US stock market and EPU have almost the same lead-lag structure as that of the UK. This result is not unexpected, since EPU and the stock markets of the two countries bear a strong resemblance. Similar to that of the the UK, the US lead-lag relationship has large fluctuations during the 2008 financial crisis and the 2020 COVID-19 pandemic. The $\left\langle x(t)\right\rangle$ reached its maximum and minimum values on August 5, 2020 (16.39) and August 4, 2008 (-83.61), respectively. Between these two periods, the US lead-lag path reveals that the stock market leads policy uncertainty slightly. 

\begin{table}[!ht]\addtolength{\tabcolsep}{2pt}
	\footnotesize
	\caption{Summary of the $\left\langle x(t)\right\rangle $ between EPU and stock market index.}
	\label{Tb:xav:Statistical}
	\medskip
	\centering
	\begin{threeparttable}
		\begin{tabular}{*{9}{c}}
			\hline\hline
			$\left\langle x(t)\right\rangle $	& Length & Mean & Median & Max & Min & Positive Values\%  & Negative Values\%   \\
			\hline						
			CNEPU-SSEC&3393&37.31&35.16&68.87&7.93&100.00&0.00\\
			UKEPU-FTSE100&3517&-2.21&-1.27&47.36&-57.48&30.85&69.15\\
			USEPU-S\&P500&3515&-5.38&-3.35&40.88&-67.17&16.39&83.61\\		
			\hline\hline		
		\end{tabular}
	\end{threeparttable}
\end{table}

As Table~\ref{Tb:xav:Statistical} reports, the mean and median of the US $\left\langle x(t)\right\rangle $ is -5.38 and -3.35 respectively. The negative path accounts for 83.61\% of the whole paths. These results suggest that the US stock market index plays a leading role in general, whilst EPU takes the lead occasionally. Our findings are similar to the earlier results of \cite{Zhou-Sornette-2006-JMe}, who find the causal arrow flowing from the US stock market to the treasury yields.  

To sum up, dynamic correlation between economic policy uncertainty and stock market returns is vulnerable to international shocks like the 2008 financial crisis. This finding is in line with the work of \cite{antonakakis2013dynamic}, \cite{arouri2016economic} and \cite{xiong2018time}.

\section{Conclusion}
\label{S5:conclusion}
We explore the time-varying co-movements between local economic policy uncertainty and stock market returns using the TOPS method. In China EPU holds an overwhelming position during the entire sample period, while in the UK, the stock market leads EPU slightly during the whole period. We observe almost the same lead-lag relationship in the US as in the UK. Moreover, extreme volatility of the interaction between EPU and domestic stock markets corresponds to crash periods like the financial crisis, the 2015 stock market crash and the COVID-19 pandemic. 

We conclude that the Chinese stock market is different from developed markets. It is still highly policy-driven, which makes policy-based prediction possible. Domestic EPU deserves more attention from Chinese stock investors and traders as it is still crucial for them to understand market behavior. They can use the information in EPU to forecast the development of Chinese stocks and adjust their portfolios accordingly. 
Besides, much needs to be done by policy makers to improve efficiency and mechanisms of China stock market.  Compared with China, stock markets of the UK and the US are far more efficient in reflecting information and news except during extremely turbulent periods like the 2008 financial crisis and the COVID-19 pandemic. In such special periods regulators should and would take measures to support financial markets and calm investors' sentiments, though big measures to support market confidence would 
take time to be effective. Thus it might be wise to stay in the market to catch the potential opportunities for arbitrage and above average profits. More generally, we complement the literature on causality between a country-specific EPU and stock market returns.

\section*{Acknowledgements}
This work was supported by the National Natural Science Foundation of China (12005064, 11805119).


\end{document}